\documentclass[aps,PRB,reprint,superscriptaddress,preprintnumbers]{revtex4-2}
\usepackage[T1]{fontenc}
\usepackage{times}
\usepackage{graphicx,color}
\usepackage{amsfonts,amsmath,amssymb,amsbsy}
\usepackage[colorlinks=true, linkcolor=blue, citecolor=blue, urlcolor=blue]{hyperref}
\usepackage{float}

\let\mathbf=\boldsymbol
\def\red#1{\textcolor{red}{#1}}

\let\emph=\red

\begin{document}

\title{Qubits based on merons in magnetic nanodisks}

\author{Jing Xia}
\affiliation{Department of Electrical and Computer Engineering, Shinshu University,
Wakasato 4-17-1, Nagano 380-8553, Japan}
\author{Xichao Zhang}
\affiliation{Department of Electrical and Computer Engineering, Shinshu University,
Wakasato 4-17-1, Nagano 380-8553, Japan}
\author{Xiaoxi Liu}
\affiliation{Department of Electrical and Computer Engineering, Shinshu University,
Wakasato 4-17-1, Nagano 380-8553, Japan}
\author{Yan Zhou}
\affiliation{School of Science and Engineering, The Chinese University of Hong Kong,
Shenzhen, Guangdong 518172, China}
\author{Motohiko Ezawa}
\affiliation{Department of Applied Physics, The University of Tokyo, 7-3-1 Hongo, Tokyo
113-8656, Japan}

\begin{abstract}
Merons and skyrmions are classical topological solitons. However, they will become quantum mechanical objects when their sizes are of the order of nanometers. Recently, quantum computation based on nanoscale skyrmions was proposed.
Here, we propose to use a nanoscale meron in a magnetic nanodisk as a qubit, where the up and down directions of the core spin are assigned to be the qubit states $|0\rangle$ and $|1\rangle$. 
First, we show numerically that a meron with the radius containing only $7$ spins can be stabilized in a ferromagnetic nanodisk classically.
Then, we show theoretically that universal quantum computation is possible based on merons by explicitly constructing the arbitrary phase-shift gate, the Hadamard gate, and the CNOT gate. They are executed by magnetic field or electric current.  
It would serve as a qubit with long coherence time as a remnant of topological stability from its classical counterpart.
\end{abstract}

\date{\today }
\maketitle






\bigskip\noindent{\usefont{T1}{phv}{b}{n}{\large Introduction}}

Quantum computation is carried out with the use of quantum mechanical states~\cite{Feynman,DiVi,Nielsen}, where superpositions and entanglements play 
essential roles. In order to execute arbitrary quantum algorithms, it is
enough to construct the $\pi /4$ phase-shift gate, the Hadamard gate, and the
controlled NOT-(CNOT) gate. Namely,  universal quantum computation is made
possible by these three gates according to the Solovay-Kitaev theorem~\cite{Deutsch,Dawson,Universal}.

A fundamental problem is how to construct qubits with the use of actual
materials. Attempts have been made in superconductors~\cite{Nakamura},
photonic systems~\cite{Knill}, quantum dots~\cite{Loss}, trapped ions~\cite{Cirac}, and nuclear magnetic resonance~\cite{Vander,Kane}. Recently, qubits based on nanoscale skyrmions have also been proposed~\cite{Psa,SkBit}. However, the coherence time is yet to be improved.

The simplest example of a qubit is a single spin, where the up spin is
assigned to the quantum state $|0\rangle$ and the down spin is assigned to
the state $|1\rangle$. The one-qubit gate operation is executed by applying
magnetic field, where the Larmor precession changes the direction of the
spin. The Heisenberg interaction gives a two-qubit gate operation~\cite{Loss,Kane}. However,
the problem is that the coherence time is too short in order to execute
quantum algorithms.

\begin{figure}[t]
\centerline{\includegraphics[width=0.48\textwidth]{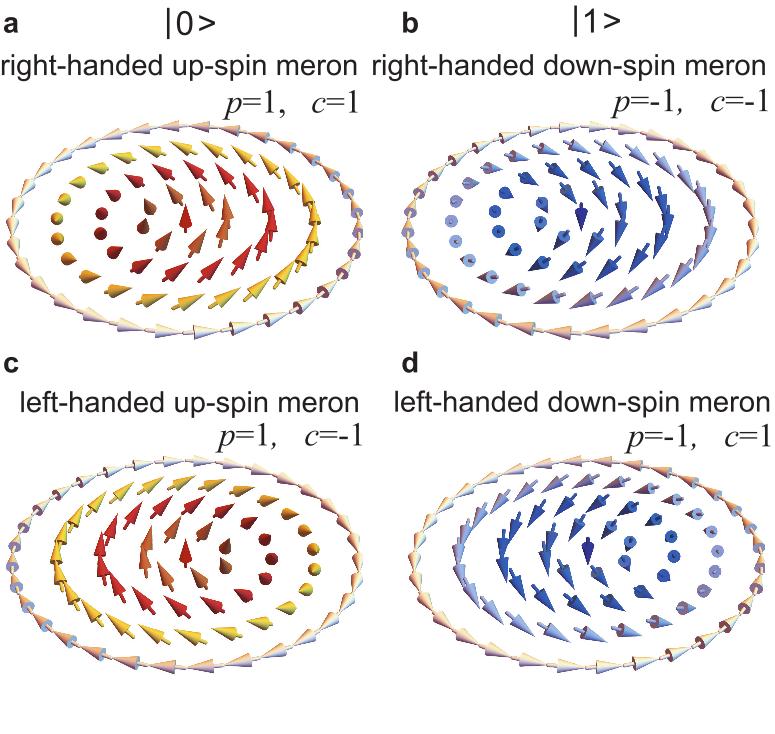}}
\caption{Illustration of Bloch-type merons with \textbf{a} polarity $p=1$ and chirality $c=1$ representing the qubit state $\left\vert
0\right\rangle $, \textbf{b} $p=-1$ and $c=-1$ representing $\left\vert 1\right\rangle$, (c) $p=1$ and $c=-1$ and (d) $p=-1$ and $c=1$. \textbf{a} and \textbf{b} are the right-handed merons, while \textbf{c} and \textbf{d} are left-handed merons. The left-handed merons have higher energy than the right-handed merons due to the DMI. 
The arrow represents the spin direction. 
The out-of-plane spin components are color coded by the red and blue color scheme.}
\label{FigMeronIllust}
\end{figure}

\begin{figure*}[t]
\centerline{\includegraphics[width=0.88\textwidth]{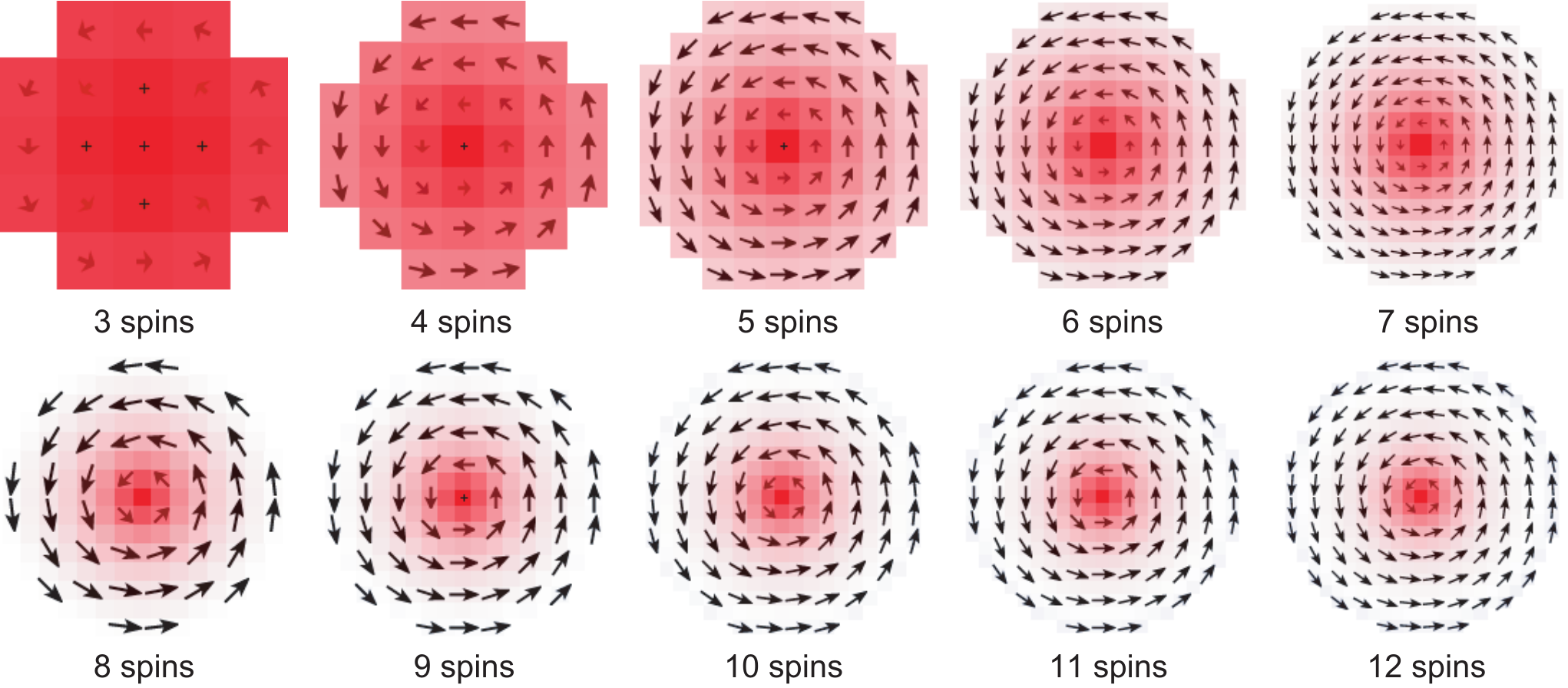}}
\caption{Simulated ground states for the meron for radius $n=3,4,\cdots, 12$. The mesh size is $0.4$nm$\times 0.4$nm. The in-plane spin direction is indicated by the arrow. The out-of-plane spin component is color coded: white is in-plane, and red is out of the plane. We have used the material parameters for MnSi~\cite{Toma}, where $A_{\text{ex}}=0.32$pJ/m, $D=0.115$mJ/m$^2$, $M_s=152$kA/m and $K=-0.5$MJ/m$^3$. See Eq.(\ref{Hamil}).}
\label{FigSquareMeron}
\end{figure*}
To overcome this problem, we focus on the core spin in a  nanoscale disk made of a chiral ferromagnet. Here, the magnetic dipole-dipole
interaction (DDI) and the easy-plane magnetic anisotropy force the spin direction to make a clockwise or anticlockwise circular rotation in the disk plane, forming a vortex-like structure called a meron~\cite{Kikuchi,Yamada,Bohl,Hertel,Nakano,Curcic,Goto,Ono_NC, Uhlir,Wintz,Sira}, as
illustrated in Fig.~\ref{FigMeronIllust}. It is a ground-state texture in disk geometry. The direction of the spin
circulation is called the chirality.
Besides, the direction of the core spin, which points upward or downward, is called the polarity.
Therefore, there are four types of merons
depending on the polarity and chirality, as illustrated in Fig.~\ref{FigMeronIllust}.
However, the Dzyaloshinskii-Moriya interaction (DMI) in chiral ferromagnets correlates the polarity
and the chirality~\cite{Ono_NC}. Hence, only right-handed merons shown in Figs.~\ref{FigMeronIllust}\textbf{a} and \textbf{b} are degenerated ground states~\cite{Ono_NC}, which serve as
a classical bit. The meron structure is topologically protected when the
sample is infinitely large. However, when its size is of the order of $100$ nm, it is quite stable and yet it is possible to reverse the core spin. The
core-spin direction could be reversed and read out by magnetic field~\cite{Hertel,Kikuchi,Curcic} or electric current~\cite{Yamada,Bohl,Nakano}.
Indeed, a random-access memory has been realized experimentally based on merons~\cite{Bohl}.

\begin{figure*}[t]
\centerline{\includegraphics[width=0.99\textwidth]{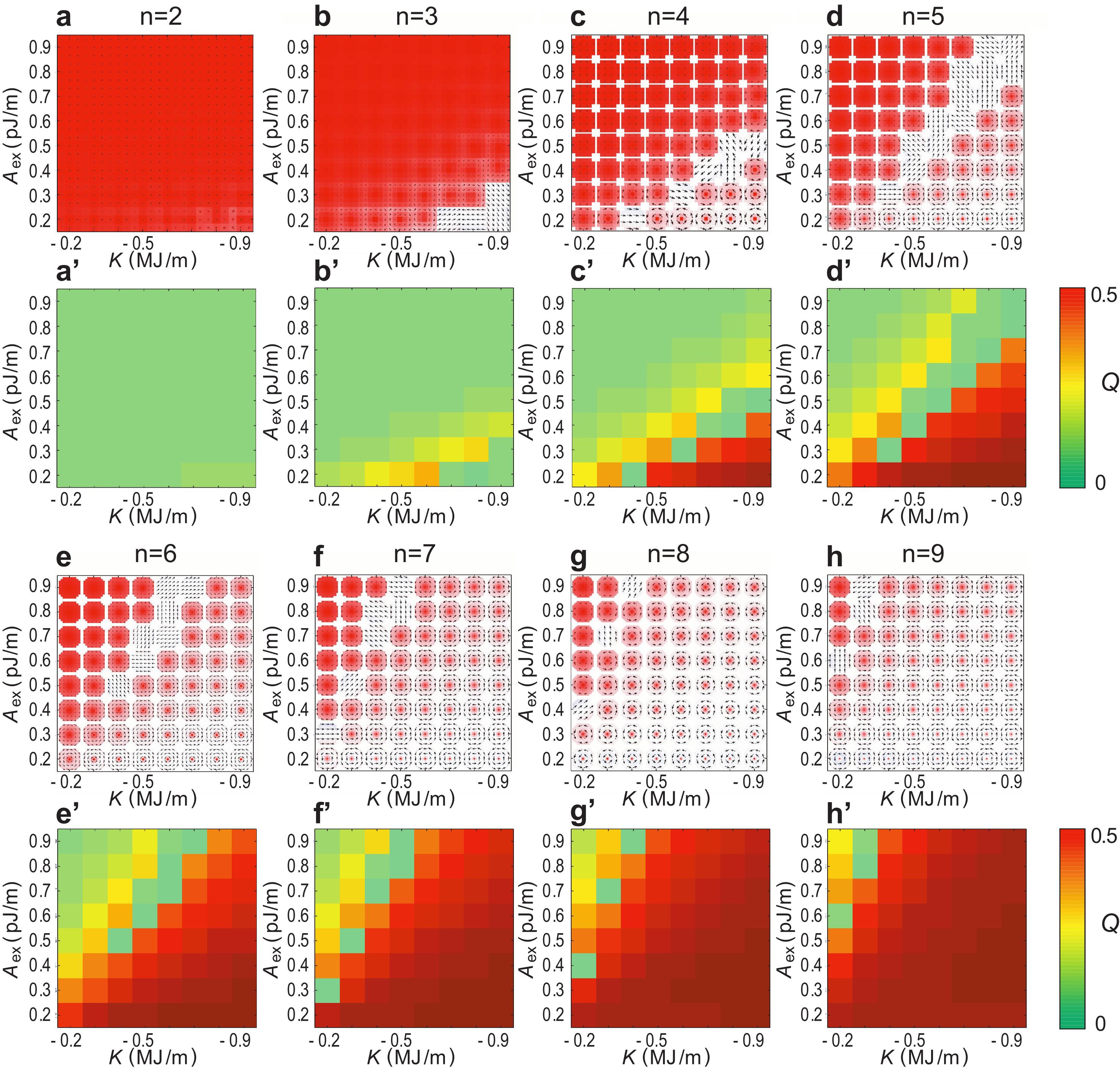}}
\caption{\textbf{a}$\sim$\textbf{h} Stability diagram of a nanoscale meron with radius $n$ for various exchange interaction (vertical axis) and easy-plane magnetic anisotropy (horizontal axis), where $n=2,3,\cdots,9$. \textbf{a}'$\sim$\textbf{h}'  Pontryagin number $Q$ of the corresponding meron together with the color scale for $Q$. We note that the material parameters for MnSi~\cite{Toma} is given by  $D=0.115$mJ/m$^2$ and $M_s=152$kA/m,}
\label{FigPhaseAll}
\end{figure*}

In this work, we propose to use a nanoscale meron in a magnetic nanodisk as a qubit, where it simulates a single
spin with a longer coherent time.
First, we study numerically how much the size of a classical meron can be made small. 
We find that a meron with the radius containing only $7$ spins is stable 
by assuming typical material  parameters taken from MnSi.
It is of the order of $3$ nm as the lattice constant is $0.4$ nm.
When the radius of the magnetic nanodisk is of the order of nanometers, the
quantum effect will be dominant. 
A nanoscale meron is uniquely specified by the
direction of the core spin. Thus, we assign the up-spin state as $|0\rangle $
and the down-spin state as $|1\rangle $. Their superposition is allowed
quantum mechanically, which represents the qubit. The coherence time is longer when the number of spins constituting a nanodisk is larger. 

The Zeeman effect due to the magnetic field induces the Pauli Z operation to
this qubit. By controlling the time duration of the Zeeman field, it is
possible to construct an arbitrary phase-shift gate including the $\pi /4$
phase shift gate. Furthermore, by applying magnetic field or electric
current, it is possible to flip a spin, which acts as the Pauli X gate.
Sequential applications of the Pauli Z and X gates produce the Hadamard
gate. Finally, the Ising interaction between layered merons produces the
controlled-Z (CZ) gate. Sequential application of the CZ and Pauli Z gates
produces the CNOT gate.

\bigskip\noindent{\usefont{T1}{phv}{b}{n}{\large Results}}

\textbf{Classical meron in a frustrated magnet.}
A meron is formed when the spin system has a nanoscale disk geometry. 
It is a vortex-like circulating structure of spins, 
where the spins on the circumference lie within the plane while the core spin points upward or downward,
forming the Bloch structure due to the DDI, as illustrated in Fig.~\ref{FigMeronIllust}. 

A meron is specified by the spin-circulation direction called the chirality $c=\pm 1$
and the core-spin direction called the polarity $p=\pm 1$.
Here, $c=1$ ($c=-1$) for the anti-clockwise (clockwise) rotation,
and $p=1$ ($p=-1$) for the up (down) spin.

The spin texture located at the coordinate center is parametrized as%
\begin{equation}
\mathbf{m}\left( x,y\right) =(\sin \theta (r)\cos \phi ,\sin \theta (r)\sin
\phi ,\cos \theta (r)),  \label{mr}
\end{equation}%
with
\begin{equation}
\phi =\varphi +\eta +\pi /2,  \label{Varphi}
\end{equation}%
where $\varphi $ is the azimuthal angle ($0\leq \varphi <2\pi $) satisyfing 
$x=r\cos \varphi $, $y=r\sin \varphi $. We note that there is a difference
from the conventional definition in Eq.~(\ref{Varphi}) by the angle $\pi /2$, where
$\eta =0$ corresponds to $c=1$, and $\eta =\pi$ corresponds to $c=-1$.
The polar angle $\theta$ is subject to
\begin{equation}
\theta (0) = 0,\pi ,\qquad \lim_{r\rightarrow R } \theta (r)=\pi /2,
\end{equation}%
where $R$ is the radius of the nanodisk, while
$\theta (0) = 0$ corresponds to $p=1$ and $\theta (0) = \pi$ corresponds to $p=-1$.
The meron with $cp=1$ is called right handed and the one with $cp=-1$ is called left handed.

There are two topological numbers defining the meron. One is the skyrmion number,
\begin{equation}
Q\equiv -\frac{1}{4\pi }\int \mathbf{m}\left( \mathbf{r}\right) \cdot \left(
\partial _{x}\mathbf{m}\left( \mathbf{r}\right) \times \partial _{y}\mathbf{m%
}\left( \mathbf{r}\right) \right) d^{2}dxdy,
\end{equation}%
which is  given by $Q=p/2$ depending on the polarity $p$.
Note that $Q$ is a half integer for the meron.

The other is the winding number defined by
\begin{equation}
\omega \equiv\int \left( \mathbf{m}\times \frac{\partial \mathbf{m}}{\partial \varphi }\right) _{z}d\varphi =c,
\end{equation}%
which depends on the chirality $c$.

There are four degenerate merons with $c=\pm 1$ and $p=\pm 1$ in the absence of the DMI. However, the DMI correlates the polarity and the chirality. The DMI is induced by the inversion symmetry breaking due to the interface between the nanodisk and the substrate~\cite{Ono_NC}.
As a result, the right-handed merons are energetically favored~\cite{Ono_NC}.
We assign the merons with $p=1$ and $p=-1$ to the classical states $|0\rangle$ and $|1\rangle$, respectively.

It is a nontrivial problem how much the size of a meron can be made small. 
Let us call it a meron with radius $n$, when its radius contains $n$ spins.
We have performed simulations on the stability of a relaxed static meron with radius $n$, $n=2, 3, \cdots, 9$, by embedding it in the $(2n-1)\times (2n-1)$ square lattice, as shown in Fig.~\ref{FigSquareMeron}. 
The simulations are carried out under the framework of micromagnetics, where we include the ferromagnetic exchange, the DMI, the magnetic DDI, and the easy-plane magnetic anisotropy  (see Methods).
As a concrete instance,
we have used the material parameters for MnSi~\cite{Toma}, where $A_{\text{ex}}=0.32$pJ/m, $D=0.115$mJ/m$^2$, $M_s=152$kA/m and $K=-0.5$MJ/m$^3$. 
The simulated ground state of a meron is demonstrated in Fig.~\ref{FigSquareMeron} for $n=3,4,\cdots,12$. We find that a meron is formed for $n\geq 4$. In addition, the radius of the core marked in red is almost identical and contains only $4$ spins irrespective of the nanodisk radius $n$ as in Fig.\ref{FigSquareMeron}.

The stability diagram showing whether the ground state is a meron or a ferromagnetic state is given in Fig.~\ref{FigPhaseAll}, where the exchange energy and the easy-plane magnetic anisotropy energy are varied with the material parameters $D=0.115$mJ/m$^2$ and $M_s=152$kA/m being fixed to those of MnSi~\cite{Toma}.
The formation of a meron is confirmed by the spin texture as in Figs.~\ref{FigPhaseAll}\textbf{a}$\sim$\textbf{h} and by the Pontryagin number as in Figs.~\ref{FigPhaseAll}\textbf{a}'$\sim$\textbf{h}'.

There are two features. 
One is that a small size meron is stabilized for a small value of the exchange interaction. It is understood that the exchange interaction becomes large for a large spin angle between the adjacent spins. 
The spin angle becomes large and the small exchange interaction has an advantage for a nanoscale meron. The other feature is that the large easy-plane magnetic anisotropy stabilizes a nanoscale meron. It is natural because the meron has an in-plane vortex structure except for the core. The radius can be as small as $7$ spins. The requirement of the exchange interaction and the easy-plane magnetic anisotropy is relaxed for a larger size of a meron.

\textbf{Control of a meron core spin.} 
If we apply an external magnetic field along the $z$ axis, the Zeeman effect splits the energy between the up and down spins.
\begin{equation}
H_{B_z}=\alpha _{B_z}B_z\sigma_z =\alpha _{B_z}B_z(|0\rangle\langle 0|-|1\rangle\langle 1|),
\end{equation}%
where $\alpha _{B_z}$ is a constant.

If we apply an external magnetic field along the $x$ axis,  where the effective Hamiltonian for the core spin is represented as
\begin{equation}
H_{B_x}=\alpha _{B_x}B_x\sigma_x =\alpha _{B_x}B_x(|0\rangle\langle 1|+|1\rangle\langle 0|),
\end{equation}%
where $\alpha _{B_x}$ is a constant. The flip of the spin is also induced by applying electric current.

\begin{figure}[t]
\centerline{\includegraphics[width=0.48\textwidth]{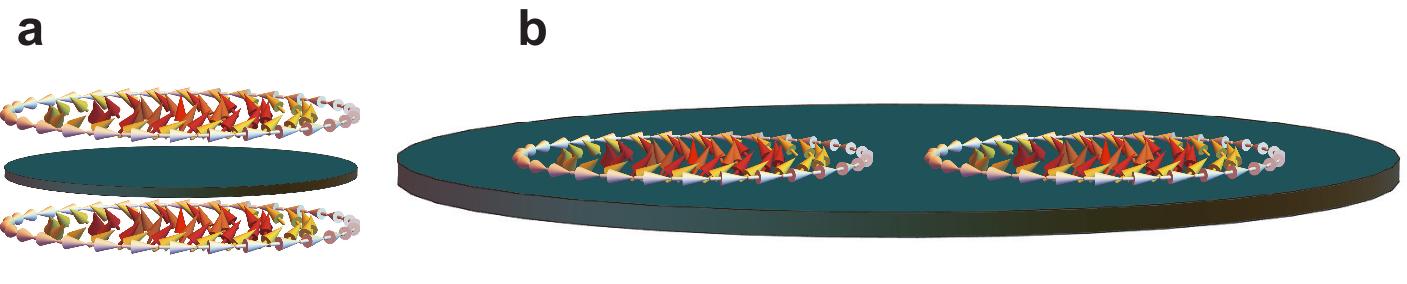}}
\caption{Illustration for the Ising interaction between two merons. \textbf{a} Vertical configuration and \textbf{b} horizontal configuration. Dark green cylinders represent a spacer in \textbf{a} and a substrate in \textbf{b}.}
\label{FigMeronLayer}
\end{figure}

We consider a bilayer nanodisk, where two nanodisks are placed vertically (Fig.~\ref{FigMeronLayer}\textbf{a}). The exchange interaction between two spins reads
\begin{equation}
H_{\text{Ising}}=J_{\text{exchange}}\sigma_z^{(1)}\otimes \sigma_z^{(2)}.
\end{equation}%
Another mechanism is to use the DDI, where the two nanodisks are placed vertically or horizontally (Figs.~\ref{FigMeronLayer}\textbf{a} and \textbf{b}). It also produces the Ising interaction because the core spin direction is fixed to be up or down.

\textbf{Core-spin qubit.} 
We focus on  the right-handed merons, which have lower energy than the left-handed merons in the presence of the DMI. 
We consider a nanodisk of the order of nanometers, where a superposition of  the up and down spins is a 
quantum mechanical state. In this regime, the up and down states of the core spin may act as a qubit. 
We assign the meron with the up core-spin ($p=1$) as the quantum state $|0\rangle$ and the one with the down core-spin ($p=-1$) as the quantum state $|1\rangle$, as illustrated in Fig.~\ref{FigMeronIllust}.

The phase-shift gate is constructed with the use of $H_{B_z}$. The Hadamard gate is constructed with the use of $H_{B_z}$ and $H_{B_x}$. The CNOT gate is constructed with the by $H_{\text{Ising}}$ and $H_{B_z}$. See Methods for details.

\textbf{Coherence time.}
The skyrmion-number conservation prohibits the core spin to flip when the sample is infinitely large. Then, the coherence time is infinite.
This is the topological protection.
Physically, it follows from the fact that it costs infinitely large energy to inverse spin directions in an infinitely large sample.
The topological protection is lost when the sample size is small. Indeed, when 
its size is of the order of $100$ nm, it is possible to flip the core spin by applying magnetic field~\cite{Hertel,Kikuchi,Curcic} or electric current~\cite{Yamada,Bohl,Nakano}. 

\begin{figure*}[t]
\centerline{\includegraphics[width=0.98\textwidth]{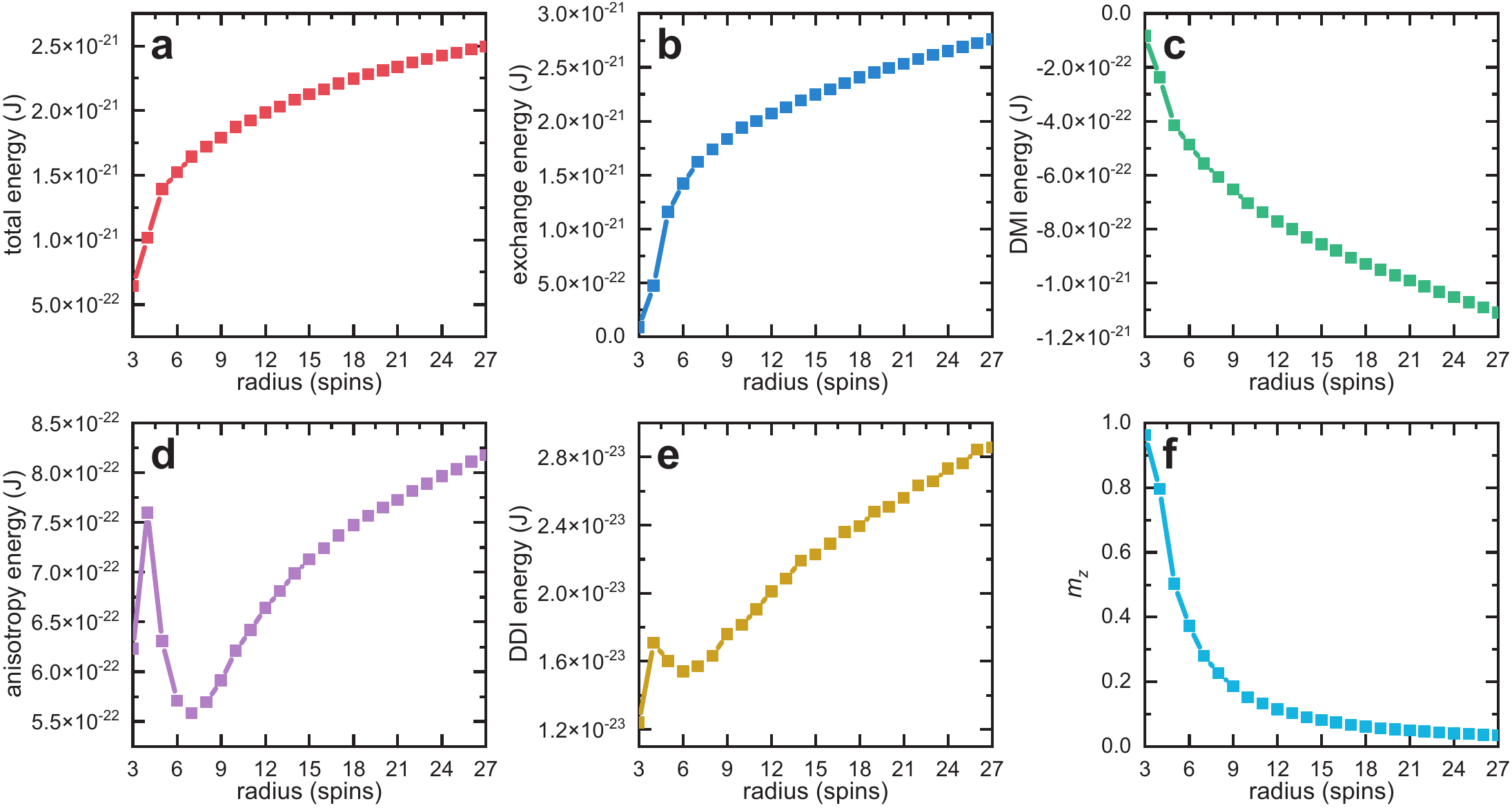}}
\caption{Size dependence of the energy of the meron in units of the Joule. \textbf{a} The total energy.  \textbf{b} The exchange energy. \textbf{c} The DMI energy. \textbf{d} The easy-plane magnetic anisotropy energy. \textbf{e} The magnetic dipole-dipole energy.  \textbf{f} Mean magnetization. 
The parameters are the same as in Fig.\ref{FigSquareMeron}.}
\label{FigEnergy}
\end{figure*}

The two merons representing $|0\rangle$ and $|1\rangle$ are obstructed by an energy barrier made of the exchange energy, easy-plane magnetic anisotropy and the DDI. We make an estimation for a small size classical meron based on the energy (\ref{Hamil}) in Methods.
The size dependence of various energies including the total, exchange, DMI, easy-plane magnetic anisotropy and magnetic DDI energies is shown in Fig.~\ref{FigEnergy}. The total energy increases as the increase of the meron size as shown in Fig.~\ref{FigEnergy}\textbf{a}. The total energy is mainly determined by the exchange energy as shown in Fig.~\ref{FigEnergy}\textbf{b}. Roughly speaking, the coherence time is proportional to the total energy because it is necessary to overcome the total energy to flip the core spin.
Thus, it is a dynamical problem to optimize the radius of a meron to make the coherent time long enough without loosing the quantum mechanical property. 

There are several features in the size dependence of the energy. First, the DMI decreases as the increase of the meron size as shown in Fig.~\ref{FigEnergy}\textbf{c}. It is understood as follows. The DMI is proportional to the spin angles between the adjacent sites. The spin angle is small for larger size merons because the spin texture becomes smooth. As a result, the DMI energy is smaller for larger size merons.
Second, there are cusp structure in the magnetic anisotropy energy and the DDI energy for merons with $n\leq 6$ as shown in Figs.~\ref{FigEnergy}\textbf{d} and \textbf{e}. They correspond to the fact that a meron is not formed but the ground state is a ferromagnetic state. 

The mean magnetization becomes smaller for larger-size merons as shown in Fig.~\ref{FigEnergy}\textbf{f}. It means that the size of the core spin is almost identical and the total spin texture looks more like a vortex structure for larger-size merons.

\textbf{Initialization.} We apply magnetic field to the sample and raise the temperature, where the system is a paramagnet. We have numerically checked that the polarity is chosen to be up by applying small external magnetic field. When we cool down the sample, right-handed up-spin merons are nucleated. 
This is the initialization of the quantum state $|00\cdots 0\rangle$.

\textbf{Read out.}
The polarity can be observed by the full-field soft X-ray transmission microscopy~\cite{Ono_NC,Uhlir}, magnetic force microscopy~\cite{Wacho,Yamada} or magnetic tunneling junction~\cite{Kasai}. The polarity is fixed to be up or down by the observation. Hence, the quantum state is fixed to be $|s_1s_2\cdots s_N\rangle$ with $s_j=0, 1$ as in the standard quantum computation.

\bigskip\noindent{\usefont{T1}{phv}{b}{n}{\large Discussions}}

The numerical estimation suggests that the minimum size of a meron is the order of $3$ nm. The energy barrier to flip the polarity is of the order of $10^{-21}$J, which is of the order of 70K. It is necessary to cool down the temperature lower than 70K to create such a meron.

We have argued that a nanoscale meron acts as a qubit and that universal quantum computation is possible.
It would behave like a single spin with a longer coherence time, being supported by the meron structure. 
It may solve the problem of short coherence time in qubits.

In the skyrmion-based quantum computations, frustrated magnets are used~\cite{Psa,SkBit}. On the other hand, an ordinary ferromagnet is enough for the present proposal. It is a merit because there are plenty of ferromagnets compared to frustrated magnets hosting nanoscale skyrmions. 

So far, we have discussed to use a meron as a qubit. However, there are four degenerate states in the absence of the DMI. Hence, it is possible to construct a qudit possessing $|0\rangle \equiv |00\rangle$, $|1\rangle\equiv |01\rangle$, $|2\rangle\equiv |10\rangle$ and $|3\rangle\equiv |11\rangle$. It is a benefit that information is doubled compared to the qubit based on the right-handed meron. However, there are issues to be explored. First, it is not clear how to introduce the Pauli Z Hamiltonian for the chirality, where the energies of the clockwise and anticlockwise chirality merons should be made different. Furthermore, it is hard to initialize the chirality of all merons identical. Next, the Ising interaction exists in both the core spin and the spins at the circumference of the nanodisk. However, it is necessary to apply the Ising coupling gate independently to the polarity and the chirality in order to execute universal quantum computation. They are future problems.  

\bigskip\noindent{\usefont{T1}{phv}{b}{n}{\large Methods}}

\textbf{Simulations of spin configurations.}
The static spin configurations in the magnetic nanodisk are simulated by using the GPU-accelerated micromagnetic simulator MUMAX3 developed at Ghent University~\cite{MUMAX}.
The energy density of the system is given by
\begin{align}
E&=-A_{\text{ex}}\sum_{\langle i,j\rangle }\mathbf{m}_i\cdot \mathbf{m}_j - \sum_i K(\mathbf{m}_i\cdot \mathbf{e}_{z})^{2}\notag\\
&-\mu_0M_s\sum_i \mathbf{m}_i\cdot \frac{1}{2}\mathbf{H}_{dd}-\sum_{\langle i,j\rangle }\mathbf{D}_{ij}\cdot ( \mathbf{m}_i \times \mathbf{m}_j), \label{Hamil}
\end{align}
where $\mathbf{m}_i$ represents the local magnetic moment orientation (\textit{i.e.}, $|\mathbf{m}_i|=1$), and $A_{\text{ex}}$ represents the ferromagnetic exchange constant.
$K$ is the easy-plane anisotropy constant, which is a negative number. The axis direction $\mathbf{e}_z$ is the normal to the easy plane.
$M_s$ represents the saturation magnetization. $\mathbf{H}_{dd}$ is the magnetic-dipole-dipole interaction. 
The last term represents the bulk DMI with $\mathbf{D}_{ij}$  being the DM vector, which stabilizes Bloch-type merons.
  
The square lattice with a circular shape is used for simulations with the lattice constant being $0.4$ nm.
Open boundary conditions are used for all sample edges.
The nanodisk is assumed to be a $1$-nm-thick MnSi with bulk DMI.
The following material parameters are adopted \cite{Toma}: $A_{\text{ex}} = 0.1$~pJ/m, $\alpha = 0.3$, $M_s = 580$ kA/m, $D = 0.115$~mJ/m$^2$ and $K = 0.1$~mJ/m$^3$.  We have demonstrated that the main conclusion of this work holds for a wide range of sample sizes as in Fig.\ref{FigPhaseAll}.

\textbf{Construction of quantum gates.}
The Schr\"{o}dinger equation for qubits is
\begin{equation}
i\hbar \frac{d}{dt}\left\vert \psi \right\rangle =H\left\vert \psi
\right\rangle, 
\end{equation}%
with the Hamiltonian
\begin{equation}
H=\alpha _{B_z}B_z\sigma_z +\alpha _{B_x}B_x\sigma_x
\end{equation}%
for single qubit, and 
\begin{equation}
H_{\text{Ising}}=J_{\text{exchange}}\sigma_z^{(1)}\otimes \sigma_z^{(2)}
\label{Ising}
\end{equation}%
for two qubits. 
We control the coefficient $B_z$, $B_x$ and $J_{\text{exchange}}$ temporally. 

We first discuss single-qubit gates.
We set $B_x=0$ and 
\begin{equation}
\alpha _{B}B_{z}\left( t\right) =\hbar \theta /2t_{0}
\end{equation}%
for $0\leq t\leq t_{0}$ and $B_{z}\left( t\right) =0$ otherwise. The
solution of the Schr\"{o}dinger equation reads%
\begin{eqnarray}
U_{Z}\left( \theta \right) &=&\exp \left[ -\frac{i}{\hbar }\sigma
_{z}\int_{0}^{t_{0}}\alpha _{B_z}B_{z}\left( t\right) dt\right]  \notag \\
&=&\exp \left[ -\frac{i\theta }{2}\sigma _{z}\right] .  \label{Zrotation}
\end{eqnarray}%
This is the $z$ rotation gate by the angle $\theta $.
It gives an arbitrary phase-shift gate.
\begin{equation}
U_{\theta}=e^{i\theta /2}U_{Z}\left( -\theta\right) ,
\end{equation}%

In the similar way, we set $B_z=0$ and 
\begin{equation}
\alpha _{B_x}B_x =\hbar \theta /2t_{0}
\end{equation}%
for $0\leq t\leq t_{0}$ and $B_{z}\left( t\right) =0$ otherwise. The
solution of the Schr\"{o}dinger equation reads%
\begin{eqnarray}
U_{X}\left( \theta \right) &=&\exp \left[ -\frac{i}{\hbar }\sigma
_{x}\int_{0}^{t_{0}}\alpha _{B_x}B_x \left( t\right) dt\right]  \notag \\
&=&\exp \left[ -\frac{i\theta }{2}\sigma _{x}\right] .  \label{Xrotation}
\end{eqnarray}%
This is the $x$ rotation gate by the angle $\theta $.

$\pi /4$ \textbf{phase-shift gate.}
The $\pi /4$ phase-shift gate is realized by the $z$ rotation (\ref{Zrotation}) by the angle $-\pi /4$ as%
\begin{equation}
U_{T}=e^{i\pi /8}U_{Z}\left( -\frac{\pi }{4}\right) ,
\end{equation}%
up to the overall phase factor $e^{i\pi /8}$.

\begin{figure}[t]
\centerline{\includegraphics[width=0.48\textwidth]{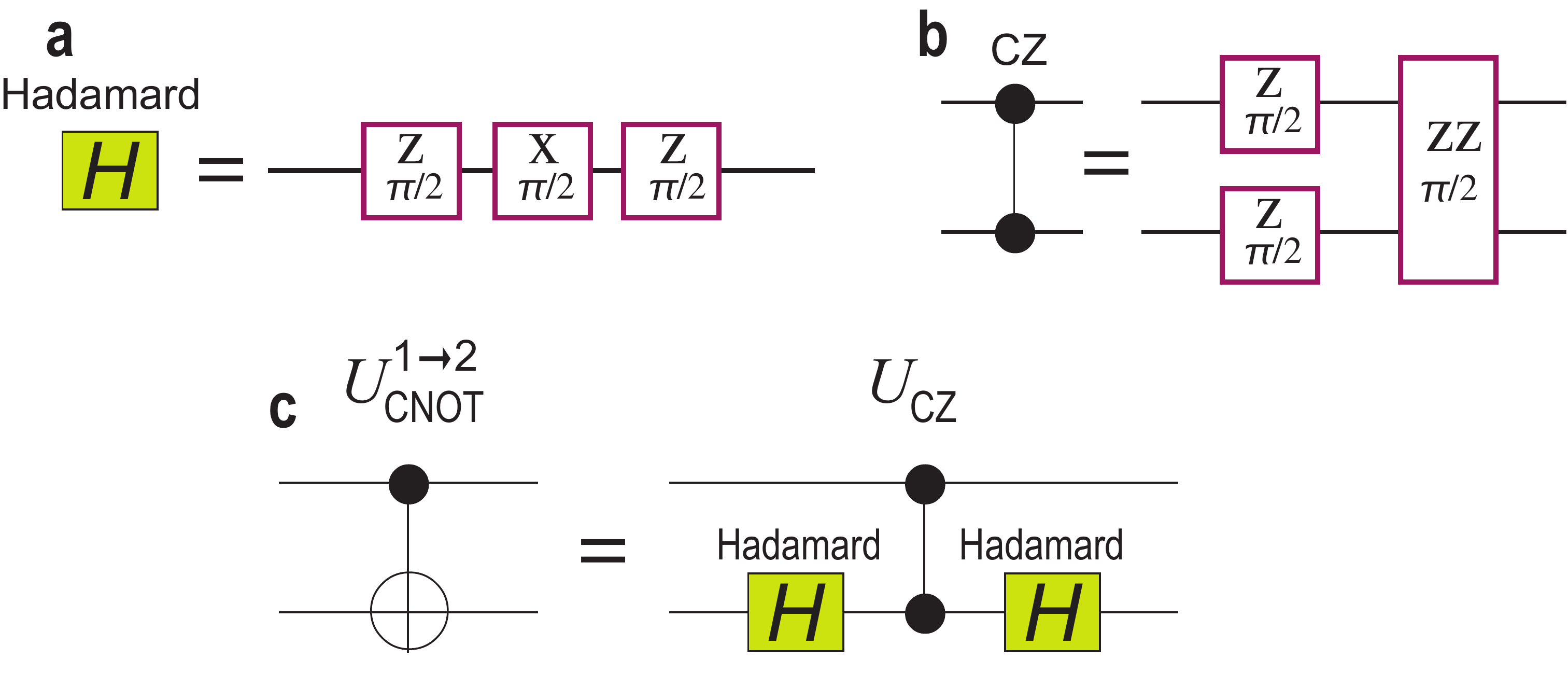}}
\caption{Quantum circuit representation of \textbf{a} the Hadamard gate in terms of
the sequential application of rotation gates as in Eq.~(\protect\ref{HZXZ})
and \textbf{b} the CZ gate in terms of the sequential applications of the $z$
rotation gate and the Ising coupling gate $U_{ZZ}$ as in Eq.~(\protect\ref{CZZZ}). \textbf{c} Quantum circuit representations of the equivalence between the CNOT
gate  as in Eq.~(\protect\ref{CNOT1}) and the CZ gate with the application of the Hadamard gates. $U_{\text{CNOT}}^{1\rightarrow 2}$.}
\label{FIgHadamard}
\end{figure}

\textbf{Hadamard gate.}
The Hadamard gate
\begin{equation}
U_{\text{H}}\equiv \frac{1}{\sqrt{2}}\left( 
\begin{array}{cc}
1 & 1 \\ 
1 & -1%
\end{array}%
\right) .
\end{equation}
is realized by a
sequential application of the $z$ rotation and the $x$ rotation~\cite{Schuch}
as%
\begin{equation}
U_{\text{H}}=-iU_{Z}\left( \frac{\pi }{2}\right) U_{X}\left( \frac{\pi }{2}\right) U_{Z}\left( \frac{\pi }{2}\right) ,  \label{HZXZ}
\end{equation}%
with the use of Eq.~(\ref{Zrotation}) and Eq.~(\ref{Xrotation}). The quantum
circuit representation of Eq.~(\ref{HZXZ}) is shown in Fig.~\ref{FIgHadamard}\textbf{a}.

Next, we discuss the two-qubits gate.
We manually control $d_m$ as a function
of time. Then, the time evolution is given by%
\begin{equation}
U=\exp \left[ -\frac{i}{\hbar }\int_{0}^{t_{0}}H_{\text{Ising}}\left(
d_{m}(t)\right) dt\right]
\end{equation}%
with $d_{m}=d_{m}(t)$ in Eq.~(\ref{Ising}). If we set $J_{\text{int}}\left(
d_{m}(t)\right) =\hbar \theta /2t_{0}$ for $0\leq t\leq t_{0}$ and $J_{\text{int}}
\left( d_{m}(t)\right) =0$ otherwise, we obtain the Ising coupling gate%
\begin{equation}
U_{ZZ}\left( \theta \right) \equiv \exp \left[ -\frac{i\theta }{2}\sigma _{z}^{\left( 1\right) }\sigma _{z}^{\left( 2\right) }\right] ,
\end{equation}%
acting on the 2-qubit in the neighboring layers.

The controlled-Z (CZ) gate $U_{\text{CZ}}$
is a unitary operation acting on
two adjacent qubits defined by
\begin{equation}
U_{\text{CZ}}=\text{diag.}(1,1,1,-1)
\end{equation}%
and constructed as~\cite{Mak}%
\begin{equation}
U_{\text{CZ}}=e^{i\pi /4}U_{Z}^{\left( 1\right) }\left( \frac{\pi }{2}\right) U_{Z}^{(2)}\left( \frac{\pi }{2}\right) U_{ZZ}^{\left( 1\right)
}\left( -\frac{\pi }{2}\right) ,  \label{CZZZ}
\end{equation}%
whose quantum circuit representation is shown in Fig.~\ref{FIgHadamard}\textbf{b}.

The CNOT gate $U_{\text{CNOT}}^{1\rightarrow 2}$
\begin{equation}
U_{\text{CNOT}}^{1\rightarrow 2}\equiv \left( 
\begin{array}{cccc}
1 & 0 & 0 & 0 \\ 
0 & 1 & 0 & 0 \\ 
0 & 0 & 0 & 1 \\ 
0 & 0 & 1 & 0%
\end{array}%
\right) ,
\end{equation}
sequential applications of the CZ gate and the Hadamard gate as
\begin{equation}
U_{\text{CNOT}}^{1\rightarrow 2}=U_{\text{H}}^{\left( 2\right) }U_{\text{CZ}}U_{\text{H}}^{\left( 2\right) },  \label{CNOT1}
\end{equation}%
where the control qubit is the skyrmion in the first layer and  target
qubit is the skyrmion in the second layer. 
The corresponding quantum circuit representation is shown in Fig.~\ref{FIgHadamard}\textbf{c}.

\textbf{Code availability.} The micromagnetic simulator MuMax used in this work is
publicly accessible at https://mumax.github.io/index.html.

\textbf{Data availability.} The data that support the findings of this study are available
from the corresponding authors upon reasonable request.






\bigskip\noindent{\usefont{T1}{phv}{b}{n}{\large Acknowledgements}}

M.E. is very much grateful to N. Nagaosa for helpful discussions on the
subject. This work is supported by the Grants-in-Aid for Scientific Research
from MEXT KAKENHI (Grants No. JP18H03676). This work is also supported by
CREST, JST (Grants No. JPMJCR20T2).
J.X. was an International Research Fellow of the Japan Society for the Promotion of Science (JSPS).
J.X. was supported by JSPS KAKENHI (Grant No. JP22F22061).
X.Z. was a JSPS International Research Fellow.
X.L. acknowledges support by the Grants-in-Aid for Scientific Research from JSPS KAKENHI (Grant Nos. JP21H01364,  JP21K18872, and JP22F22061).
Y.Z. acknowledges support by Guangdong Basic and Applied
Basic Research Foundation (2021B1515120047), Guangdong Special Support
Project (Grant No. 2019BT02X030), Shenzhen Fundamental Research Fund (Grant
No. JCYJ20210324120213037), Shenzhen Peacock Group Plan (Grant No.
KQTD20180413181702403), Pearl River Recruitment Program of Talents (Grant
No. 2017GC010293), and National Natural Science Foundation of China (Grant
Nos. 11974298, 12004320, and 61961136006).

\bigskip\noindent{\usefont{T1}{phv}{b}{n}{\large Author contributions}}

M.E. conceived the idea and conducted the project.  
J.X. and X.Z. performed numerical simulations in collaboration with X.L. and Y.Z.
All authors discussed the results and wrote the manuscript.

\bigskip\noindent{\usefont{T1}{phv}{b}{n}{\large Additional information}}

\textbf{Competing financial and non-financial interests:} All authors declare no competing financial and non-financial interests.


\begin{thebibliography}{99}

\bibitem{Feynman} R. Feynman, Simulating physics with computers, Int. J. Theor. Phys. \textbf{21}, 467 (1982).

\bibitem{DiVi} D. P. DiVincenzo, Quantum Computation, Science \textbf{270}, 255 (1995).

\bibitem{Nielsen} M. Nielsen and I. Chuang, "Quantum Computation and Quantum
Information", Cambridge University Press, (2016); ISBN 978-1-107-00217-3.

\bibitem{Deutsch} D. Deutsch, Quantum theory, the Church?Turing principle and the universal quantum computer, Proceedings of the Royal Society A. \textbf{400}, 97 (1985).

\bibitem{Dawson} C. M. Dawson and M. A. Nielsen, The Solovay-Kitaev algorithm, quant-ph/arXiv:0505030.

\bibitem{Universal} M. Nielsen and I. Chuang, "Quantum Computation and
Quantum Information", Cambridge University Press, Cambridge, UK (2010).

\bibitem{Nakamura} Y. Nakamura; Yu. A. Pashkin; J. S. Tsai, Coherent control of macroscopic quantum states in a single-Cooper-pair box, Nature \textbf{398}, 786 (1999).

\bibitem{Knill} E. Knill, R. Laflamme and G. J. Milburn, A scheme for efficient quantum computation with linear optics, Nature, \textbf{409}, 46 (2001).

\bibitem{Loss} D. Loss and D. P. DiVincenzo, Quantum computation with quantum dots, Phys. Rev. A \textbf{57}, 120
(1998).

\bibitem{Cirac} J. I. Cirac and P. Zoller, Quantum Computations with Cold Trapped Ions, Phys. Rev. Lett. \textbf{74},
4091 (1995).

\bibitem{Vander} L. M.K. Vandersypen, M. Steffen, G. Breyta, C. S. Yannoni,
M. H. Sherwood, I. L. Chuang, Experimental realization of Shor's quantum factoring algorithm using nuclear magnetic resonance, Nature \textbf{414}, 883 (2001).

\bibitem{Kane} B. E. Kane, A silicon-based nuclear spin quantum computer, Nature \textbf{393}, 133 (1998).


\bibitem{Psa} C. Psaroudaki and C. Panagopoulos, Skyrmion Qubits: A New Class of Quantum Logic Elements
Based on Nanoscale Magnetization, Phys. Rev. Lett. \textbf{127}, 067201
(2021).

\bibitem{SkBit} J. Xia, X. Zhang, X. Liu, Y. Zhou, M. Ezawa, Universal quantum computation based on nanoscale skyrmion helicity qubits in frustrated magnets
arXiv:2204.04589 

\bibitem{Kikuchi} N. Kikuchi, S. Okamoto, O. Kitakami, Y. Shimada, S. G. Kim, Y. Otani and K. Fukamichi, Vertical bistable switching of spin vortex in a circular magnetic dot, J. App. Phys. 90, 6548 (2001)

\bibitem{Hertel} R. Hertel, S. Gliga,  M. Fahnle and C.M. Schneider, "Ultrafast Nanomagnetic Toggle Switching of Vortex Cores", Phys. Rev. Lett. 98, 117201 (2007)

\bibitem{Curcic} M. Curcic,  B. Van Waeyenberge,  A. Vansteenkiste,  M. Weigand,  V. Sackmann, H. Stoll, M. Fahnle, T. Tyliszczak,  G. Woltersdorf,  C. H. Back and G. Schutz, "Polarization Selective Magnetic Vortex Dynamics andCoreReversalin Rotating Magnetic Fields" Phys. Rev. Lett.  101, 197204 (2008)


\bibitem{Yamada} K. Yamada, S. Kasai, Y. Nakatani, K. Kobayashi, H. Kohno, A. Thiaville and T. Ono, Electrical switching of the vortex core in a magnetic disk, Nat. Mat. 6, 270 (2007)

\bibitem{Bohl} S. Bohlens, B. Kruger,  A. Drews,  M. Bolte,  G. Meier and
D.  Pfannkuche, Current controlled random-access memory based on magnetic
vortex handedness, Appl. Phys. Lett. 93, 142508 (2008)

\bibitem{Nakano} K. Nakano, D. Chiba, N. Ohshima, S. Kasai, T. Sato, Y. Nakatani,  K. Sekiguchi, K. Kobayashi and T. Ono, All-electrical operation of magnetic vortex core memory cell, Appl. Phys. Lett. 99, 262505 (2011)

\bibitem{Goto} M. Goto, H. Hata, A. Yamaguchi,  Y. Nakatani,  T. Yamaoka, Y. Nozaki and H. Miyajima, Electric spectroscopy of vortex states and dynamics in magnetic disks, Phys. Rev. B 84, 064406 (2011)

\bibitem{Ono_NC} M.-Y. Im, P. Fischer, K. Yamada, T. Sato, S. Kasai,
Y. Nakatani and T. Ono, Symmetry breaking in the formation of magnetic
vortex states in a permalloy nanodisk, Nat. Com. 3, 983 (2012)

\bibitem{Uhlir} V. Uhlir, M. Urbanek, L. Hladik, J. Spousta, M-Y. Im, P. Fischer, N. Eibagi, J. J. Kan,
E. E. Fullerton  and T. S. Sikola, Dynamic switching of the spin circulation in tapered magnetic nanodisks, Nat. Nanotech. 8, 341 (2013)

\bibitem{Wintz} S. Wintz,  C. Bunce, A. Neudert,  M. Korner,  T. Strache,  M. Buhl, 
A. Erbe,  S. Gemming, J. Raabe,  C. Quitmann,  and J. Fassbender, Phys. Rev. Lett.  110, 177201 (2013)

\bibitem{Sira} G. Siracusano, R. Tomasello, A. Giordano,  V. Puliafito,  B. Azzerboni,  O. Ozatay,  M. Carpentieri and G. Finocchio, Topology and Origin of Effective Spin Meron Pairs in Ferromagnetic Multilayer Elements, Phys. Rev. Lett. 117, 087204 (2016)

\bibitem{Wacho}  A. Wachowiak, J. Wiebe, M. Bode, O. Pietzsch, M. Morgenstern, and R.
Wiesendanger, Direct Observation of Internal Spin Structure of Magnetic Vortex Cores, Science 298, 577 ?2002?.

\bibitem{Kasai}  S. Kasai, K. Nakano, K. Kondou, N. Oshima, K. Kobayashi and T. Ono, Three-Terminal Device Based on the Current-Induced Magnetic Vortex Dynamics with the Magnetic Tunnel Junction
Appl. Phys. Express 1, 091302 (2008)

\bibitem{Schuch} N. Schuch and J. Seiwert, Natural two-qubit gate for quantum computation using the XY interaction, Phys. Rev. A \textbf{67}, 032301 (2003).

\bibitem{Mak} Y. Makhlin, Nonlocal properties of two-qubit gates and mixed states and optimization of quantum computations, Quant. Info. Proc. \textbf{1}, 243 (2002).

\bibitem{Toma} R. Tomasello, E. Martinez, R. Zivieri, L. Torres, M. Carpentieri and G. Finocchio, A strategy for the design of skyrmion racetrack memories, Scientific Reports \textbf{4}, 6784 (2014).

\bibitem{MUMAX} A. Vansteenkiste, J. Leliaert, M. Dvornik, M. Helsen, F. Garcia-Sanchez, and B. Van Waeyenberge, The design and verification of MuMax3, AIP Adv. 4, 107133 (2014).



\end{thebibliography}
\end{document}